\documentclass[twocolumn]{aastex701}

\usepackage{placeins}
\usepackage{enumitem}
\usepackage{amsmath, amssymb}
\usepackage{graphicx}

\begin{document}
\title{A Search for High Frequency Oscillations in TESS Cycle 7}

\author{Guanda Huang}
\affiliation{CAS Key Laboratory of Optical Astronomy, National Astronomical Observatories, Chinese Academy of Sciences, Beijing 100101, People's Republic of China;}
\affiliation{School of Astronomy and Space Science, University of the Chinese Academy of Sciences, Beijing 100049, People's Republic of China}
\email{huanggd@bao.ac.cn}

\author{Xiaodian Chen}
\affiliation{CAS Key Laboratory of Optical Astronomy, National Astronomical Observatories, Chinese Academy of Sciences, Beijing 100101, People's Republic of China;}
\affiliation{School of Astronomy and Space Science, University of the Chinese Academy of Sciences, Beijing 100049, People's Republic of China}
\email{chenxiaodian@nao.cas.cn}

\author{Shu Wang}
\affiliation{CAS Key Laboratory of Optical Astronomy, National Astronomical Observatories, Chinese Academy of Sciences, Beijing 100101, People's Republic of China;}
\affiliation{School of Astronomy and Space Science, University of the Chinese Academy of Sciences, Beijing 100049, People's Republic of China}
\email{shuwang@nao.cas.cn}

\author{Xiaobin Zhang}
\affiliation{CAS Key Laboratory of Optical Astronomy, National Astronomical Observatories, Chinese Academy of Sciences, Beijing 100101, People's Republic of China;}
\affiliation{School of Astronomy and Space Science, University of the Chinese Academy of Sciences, Beijing 100049, People's Republic of China}
\email{xzhang@bao.ac.cn}

\author{Licai Deng}
\affiliation{CAS Key Laboratory of Optical Astronomy, National Astronomical Observatories, Chinese Academy of Sciences, Beijing 100101, People's Republic of China;}
\affiliation{School of Astronomy and Space Science, University of the Chinese Academy of Sciences, Beijing 100049, People's Republic of China}
\email{licai@bao.ac.cn}

\begin{abstract}

High-quality, short-cadence photometry from TESS enables the detection of rapid oscillators 
with unprecedented sensitivity. In this work, 
we conduct a homogeneous search for high-frequency variability using 
20-second cadence light curves from TESS Cycle 7 (Sectors 84--96). 
From $\sim 3.9\times10^{4}$ light curves, 
we compute Lomb-Scargle periodograms and select candidates exhibiting at least one significant signal with $\mathrm{FAP}\le 10^{-4}$ at frequencies $f\ge 50~\mathrm{d^{-1}}$. 
After excluding previously reported objects and performing pixel-level and light-curve vetting to mitigate contamination, 
we identify 73 rapid oscillators, including 24 pulsating white dwarfs, 31 hot subdwarfs, and 18 A-F stars. 
Using an iterative prewhitening procedure, we carry out a detailed frequency analysis for each target and derive the oscillation frequencies and amplitudes. 
We further investigate the physical origins of the detected frequency content and present statistical characterizations of the rapid-oscillator sample. 
We highlight one white dwarf and one subdwarf that exhibit clear frequency multiplets consistent with rotational splitting. 
This work enlarges the sample of rapid oscillators accessible with TESS data and provides a uniformly measured frequency-amplitude catalog, 
establishing a consistent basis for future asteroseismic and population studies. 

\end{abstract}

\keywords{\uat{Stellar oscillations}{1617} --- \uat{Light curves}{918} --- 
    \uat{Variable stars}{1761} --- \uat{White dwarf stars}{1799} --- 
    \uat{Pulsation modes}{1309} --- \uat{Pulsating variable stars}{1307} ---
    \uat{B subdwarf stars}{129} --- \uat{Delta Scuti variable stars}{371}
    }


\section{Introduction} 
Rapid oscillations with periods of seconds to tens of minutes are observed across a wide range of stellar evolutionary stages, 
so the high-frequency regime is a shared diagnostic window rather than a niche corner of stellar variability. 
Among main-sequence stars, several classes of A-F type variables, including $\delta$~Scuti and roAp stars, exhibit high-frequency pulsations. 
\citet{2011AJ....142..110M} emphasized the long-standing astrophysical utility of short-period classical pulsators, 
while \citet{2010ApJ...713L.192G} showed with early \textit{Kepler} data that 
hybrid \textit{p}- and \textit{g}-mode behavior is common enough to complicate any rigid class boundary.
\cite{Balona2019, Bedding2020, 1982MNRAS.200..807K} show that high-frequency oscillations 
above $50\ \mathrm{d}^{-1}$ can be generally found in $\delta$~Scuti stars and roAp stars. 

As main-sequence stars evolve, most of them eventually become white dwarfs, 
and white dwarfs are also found to pulsate in this frequency range of particular interest, 
with typical periods from $30\ \mathrm{s}$ to $25\ \mathrm{min}$, corresponding to frequencies from $57.6$ to $2880$ cycles per day \citep{Corsico2019}. 
White-dwarf reviews by \citet{2010A&ARv..18..471A} and \citet{2019A&ARv..27....7C} 
show why these short periods are especially valuable for constraining internal chemical layering, 
envelope thickness, and even rotation in compact remnants. 

Between the main-sequence and the white-dwarf phases lie the hot subdwarfs, which also show short-period oscillations.
Subdwarf B stars are He-burning cores with extremely thin H envelopes, formed after significant mass loss
during the late red-giant phase, when most of the H envelope was depleted;
the origin of hotter subdwarf O stars is more diverse. 
He-deficient subdwarf O stars can evolve from subdwarf B stars \citep{Stroeer2007}, and the 
He-rich ones can be produced from a late hot He-flash \citep{Bertolami2008} or through a merging event of two
white dwarfs \citep{Saio2000, Saio2002}. 

Pulsating hot subdwarf B (sdB) stars are classified into three distinct categories based on their oscillation modes, 
as proposed by \citet{Uzundag2024}: $p$-mode, $g$-mode, and the intermediate `$h$-mode'. 
The $p$-mode and $g$-mode oscillations typically present signals at frequencies 
of $\gtrsim 340\ \mathrm{d}^{-1}$ and $\lesssim 55\ \mathrm{d}^{-1}$, respectively. 
The $h$-mode refers to oscillations in the intermediate frequency range ($55$--$340\ \mathrm{d}^{-1}$); 
notably, these signals were previously often categorized as $p$-modes, 
but are now explicitly distinguished as a separate class \citep{Uzundag2024}.

As for pulsating subdwarf O stars, all known targets are
found to be rapidly oscillating with periods of a few minutes, 
consistent with \textit{p}-mode oscillating subdwarf B stars \citep{Randall2016}. 
The formation pathways of these stars, 
including stellar mergers and close-binary evolution, 
shape their internal physical structures and therefore influence their pulsational properties.
Evolutionary models of these channels \citep{2018MNRAS.476.5303S, 2021ApJ...920..110S} thus provide 
an important theoretical framework for interpreting rapid p-mode and related oscillations, 
although these studies are not observational pulsation catalogs.
\citet{2020ApJ...888...49W} further demonstrated that TESS photometry can already reach rapid variability in two pulsating pre-ELM white-dwarf candidates, 
reinforcing that sub-hour oscillators span more than one compact-star subclass. 
Taken together, these studies make the $f \ge 50\,\mathrm{d}^{-1}$ domain important 
because it links phenomenology to interior physics across white dwarfs, 
hot subdwarfs, and A-F stars. 

The observational pathway to a modern high-frequency inventory runs from 
ground-based and \textit{Kepler}-era classification work to large TESS catalogs with increasingly explicit cadence choices. 
\citet{2015AJ....149...68B} analyzed 2768 \textit{Kepler} Guest Observer targets near the instability strips in a uniform search. 
They identified 207 $\gamma$~Doradus, 84 $\delta$ Scuti, and 32 hybrid candidates, 
making clear that homogeneous processing changes the apparent class balance. 
Using two large \textit{Kepler} ensembles, \citet{2018MNRAS.476.3169B} quantified instrumental biases and showed that 
ensemble studies can isolate the most informative $\delta$ Scuti targets for later mode identification. 
In parallel, \citet{2019MNRAS.485.2380M} used Gaia DR2 luminosities for more than 15,000 \textit{Kepler} A/F stars. 
They found that 18\% of the $\delta$ Scuti stars have dominant frequencies above the \textit{Kepler} long-cadence Nyquist frequency, 
and 30\% show some super-Nyquist variability. All-sky survey growth widened the context. 
\citet{2018AJ....156..241H} released 4.7 million ATLAS candidate variables, 
\citet{2020ApJS..249...18C} classified 781,602 periodic variables in ZTF DR2, 
and \citet{2021MNRAS.508.3877G} supplied a Gaia EDR3 white-dwarf candidate catalog with about 359,000 high-confidence objects for cross-identification. 
Within the TESS era itself, \citet{2020MNRAS.493.4186J} characterized an all-sky sample of about 8400 $\delta$ Scuti stars. 
\citet{2023ApJ...946L..10B} detected $\delta$ Scuti pulsations in 36 of 89 A/F-type Pleiades members. 
\citet{2024AJ....168...83O} identified 14,156 short-period variables from 30 min full-frame light curves. 
\citet{2024ApJ...972..137G} measured variability in 103,810 TESS A--F stars from the first 26 sectors. 
\citet{2025ApJS..276...57G} found 72,505 periodic variables from the first 67 sectors of 2-minute data. 
In \cite{Sahoo2020, Baran2021b, Sahoo2023}, a series of searches for variable subdwarf B stars 
was conducted with 30-minute cadence TESS Full Frame Image (FFI) data. 
\cite{Antoci2019} used 2-minute cadence data to study 
the asteroseismology of $\delta$~Scuti and $\gamma$~Doradus stars. 
\citet{Baran2023, Baran2024} conducted systematic searches for pulsating hot subdwarf stars 
using TESS data up to Sector 46 for the southern hemisphere and 
Sector 60 for the northern hemisphere, respectively.
The cadence arithmetic is central to why these surveys are complementary: 
the Nyquist limits are about $24\,\mathrm{d}^{-1}$ for 30 min, $360\,\mathrm{d}^{-1}$ for 2-minute, and $2160\,\mathrm{d}^{-1}$ for 20-second sampling. 
A Cycle 7 search in 20-second data therefore probes a frequency space that many large FFI-based catalogs cannot fully access.

In this study, we further exploit the advantage of TESS 20-second cadence data to 
conduct a systematic search for high-frequency oscillations using the most recent Cycle 7 data. 
By requiring at least one statistically significant peak with $f \ge 50\,\mathrm{d}^{-1}$, 
we apply a single Lomb--Scargle plus iterative prewhitening workflow to build 
a reusable frequency-amplitude parameter set across white dwarfs, hot subdwarfs, and A--F high-frequency pulsators. 
The goal is not only to add newly characterized rapid variables, 
but also to place rotational splitting, intermediate-frequency multiplets, 
and other diagnostically useful structures---including Fourier modulation, which has been successfully employed to confirm binary systems \citep[e.g.,][]{Shibahashi2012, Smith2022, Otani2025}---
onto a more uniform measurement scale. 
\autoref{sec:method} describes the Cycle 7 data set and the frequency-extraction and vetting procedure. 
\autoref{sec:results} gives the summary statistics, highlights representative objects, and discusses the astrophysical implications of the expanded sample. 
\autoref{sec:discussion} shows examples of aliasing and contamination that can occur 
when the large pixel scale of TESS is applied to faint targets and optical binaries. 
\autoref{sec:summary} summarizes the main conclusions.

\section{TESS Data and Identification Method}
\label{sec:method}

We downloaded all TESS 20-second cadence light-curve data from Sectors 84 through 96 from the Barbara A. Mikulski Archive for Space Telescopes (MAST)
(\dataset[doi:10.17909/t9-st5g-3177]{\doi{10.17909/t9-st5g-3177}}),
\footnote{https://archive.stsci.edu/}. 
This includes $\sim39000$ light curves, with $\sim3000$ light curves per sector. 
In this work, we adopted \texttt{time} values, representing the barycentric Julian date (BJD) 
corrected to the TESS mission reference time ($\mathrm{BJD}-2457000$), and 
\texttt{pdcsap\_flux} values, which denote `Pre-search Data Conditioning Simple Aperture Flux', 
which are corrected for instrumental trends by the 
Science Processing Operations Center (SPOC) pipeline \citep{SPOC}.

Using the \texttt{LombScargle} implementation provided in the Python package \texttt{astropy.timeseries} \citep{VanderPlas2012, VanderPlas2015}, 
the periodograms were computed over a frequency range from ${1/27.4}\ \mathrm{d}^{-1}$ to 
$2160\ \mathrm{d}^{-1}$. 
To ensure all signal peaks would be fully resolved, 
the theoretical frequency resolution ${1/27.4}\ \mathrm{d}^{-1}$ was oversampled by a factor of six in constructing the frequency grid.
We then filtered for targets showing a significant signal at frequencies $\geq50\ \mathrm{d}^{-1}$, 
defining "significant" with a threshold of false alarm probability (FAP) $\leq 0.0001$. 
This is a relatively strict criterion compared with previous works. For example, \citet{Romero2022, Romero2025} 
used a criterion of $\mathrm{FAP}\leq 0.001$. \citet{Baran2023, Baran2024} adopted a threshold of 4.5 times the median noise level, 
which corresponds to a FAP of $14\%$, following \citet{Baran2021a}. 
We intentionally adopted this conservative threshold as a practical compromise between completeness, sample reliability, 
and the efficiency of manual vetting. 
Across all sectors, applying a more relaxed threshold of 0.001 would increase the candidate list by approximately 20\%, 
whereas fewer than about 2\% of the initial candidates survive the full vetting process and enter the final sample. 
Thus, even a moderate increase in the candidate list can translate into a substantial amount of additional manual inspection. 
To test the effect of the relaxed threshold, we used Sector 84 as a benchmark. 
Applying $\mathrm{FAP}\leq 0.001$ yielded 52 additional candidates; 
however, upon manual inspection of their light curves and periodograms, 
none were identified as genuine targets of interest, except for two previously known pulsating white dwarfs from \citet{Romero2022}. 
We therefore adopted $\mathrm{FAP}\leq 0.0001$ to retain a high-quality sample while keeping the candidate-vetting process efficient.

\begin{figure*}[htbp]
\plotone{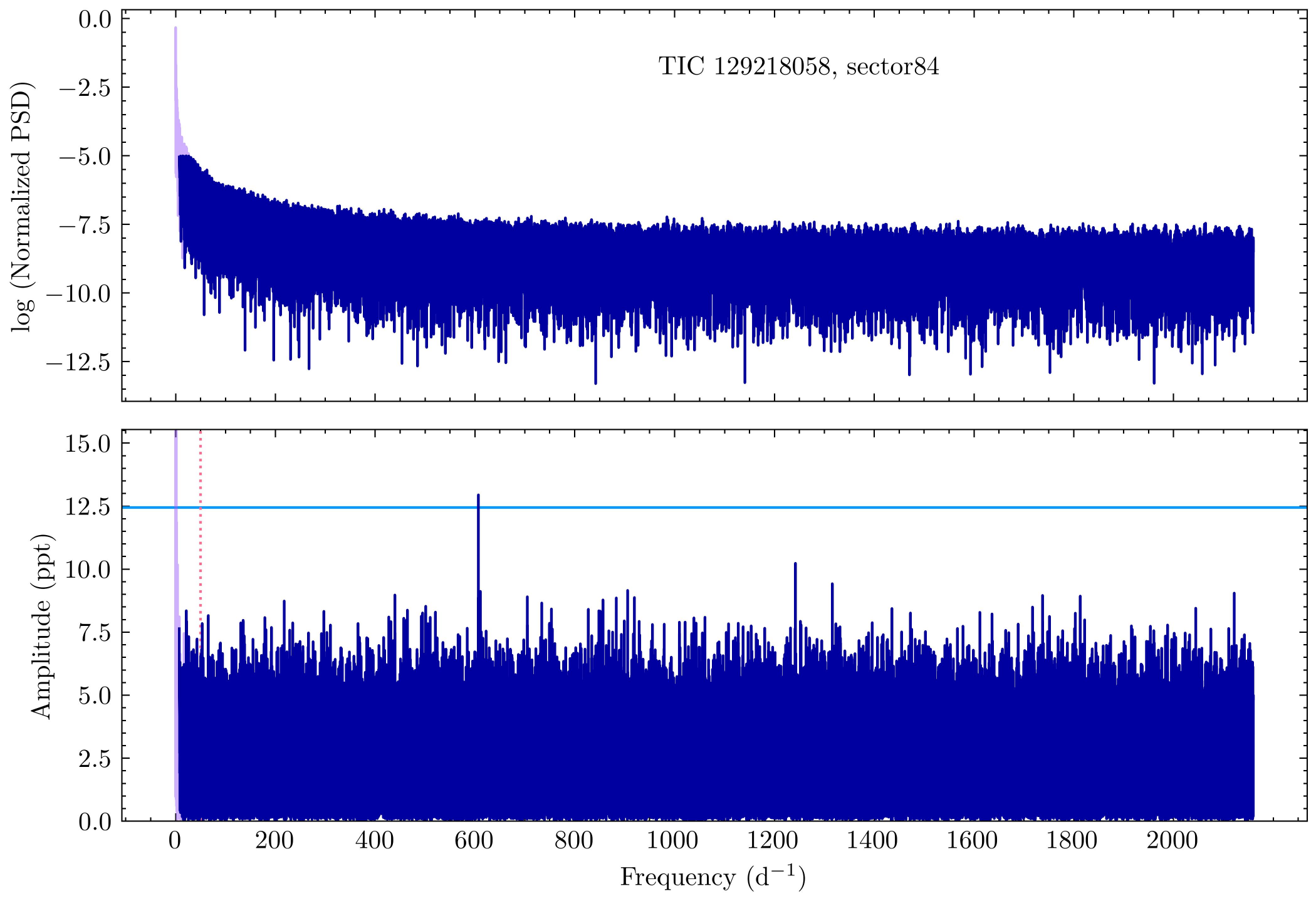} 
\caption{Periodogram analysis of the pulsating white dwarf TIC 129218058 using data from TESS Sector 84. 
Top panel: Spectral window function shown on a normalized logarithmic scale. 
Bottom panel: Amplitude periodogram. 
The blue horizontal line marks the detection threshold corresponding to 
a false alarm probability (FAP) of $10^{-4}$ (in this case $\sim 12.5\ \mathrm{ppt}$), 
while the red dotted vertical line indicates the frequency cutoff at $50\ \mathrm{d}^{-1}$. 
The purple shaded regions in both panels represent artifacts introduced by the sampling window. 
Note that the amplitude axis is truncated to highlight the genuine pulsation peak at $607.00\ \mathrm{d}^{-1}$ ($12.94\ \mathrm{ppt}$); 
the dominant spurious signal induced by sampling effects peaks at $0.57\ \mathrm{d}^{-1}$ with an amplitude of $40.25\ \mathrm{ppt}$.}
\label{fig:window function and periodogram}
\end{figure*}

Furthermore, to mitigate the effect of the sampling window function \citep{VanderPlas2018}, 
we calculated the periodogram of the window function for each target, and removed a fraction of frequencies 
most severely affected by the sampling effect. 
In practice, we removed $0.5\%$ of frequencies with the highest power in the window function periodogram, 
and this effectively removed the frequencies below $50\ \mathrm{d}^{-1}$, while minimizing the impact on the frequency range of interest.
In addition, these frequencies could be unambiguously identified as the most strongly affected by visual inspection 
of the window function periodograms (see \autoref{fig:window function and periodogram} for an example).

After filtering, we found $\sim4500$ targets with at least one signal peak in the $\geq50\ \mathrm{d}^{-1}$ region, 
with an average of $\sim 350$ candidates per sector. 
We excluded targets showing a solar-like oscillation pattern, which is commonly observed in red giants and is not the focus of this work. 
Targets showing EA-type eclipsing binary light curves were also excluded; in most of these cases, the high-frequency signals are confirmed to be harmonics of the orbital frequency. 
To focus on discovering new targets, we also excluded known targets reported in previous works by 
\cite{Romero2022, Romero2025, Balona2019, Balona2022, Uzundag2024}. 
For the remaining candidate targets, we performed a standard prewhitening procedure. 
For each target, we fit a sinusoidal signal with frequency corresponding to the largest amplitude in the spectrum, 
after which the fitted signal is subtracted from the light curve. This was repeated until no 
peak in the spectrum exceeds the detection threshold, or to a maximum of 100 iterations. 
The fitted signal is described as follows:
$$y_{fit}=\theta_0+\theta_1\sin(2\pi ft)+\theta_2\cos(2\pi ft)$$
where $f$ is the fitted frequency and $t$ is time. 

We used subdwarf candidate catalogs from \cite{Geier2019, Culpan2022} to identify our targets as subdwarfs, 
and white dwarf catalogs from \citep{GentileFusillo2018, GentileFusillo2021} to identify white dwarfs, 
since we consider these to be the most up-to-date target lists for subdwarfs and white dwarfs. 
We adopted the following standard to classify the remaining candidates as A-F stars: 
$$-0.4\leq M_G\leq4$$
$$0\leq BP-RP\leq0.8$$
where $M_G$ is Gaia absolute magnitude, calculated as:
$$M_G=m_G+5\times\log_{10}{(\varpi/1000)}+5$$
where $m_G$ is Gaia G apparent magnitude and $\varpi$ is Gaia parallax in milliarcseconds. 
After visual inspection of all periodograms, light curves, and prewhitening results of the filtered candidates, 
we present a final catalog of 73 targets (see \autoref{sec:results} and \autoref{sec:tables}).

\section{Results}
\label{sec:results}

As a result, we identify 73 targets with rapid variations at frequencies above $50\ \mathrm{d}^{-1}$, 
including 24 white dwarfs, 31 hot subdwarfs, and 18 A-F stars. The frequencies and corresponding amplitudes of these targets are listed in \autoref{sec:tables}.

\begin{figure*}[!t]
\plotone{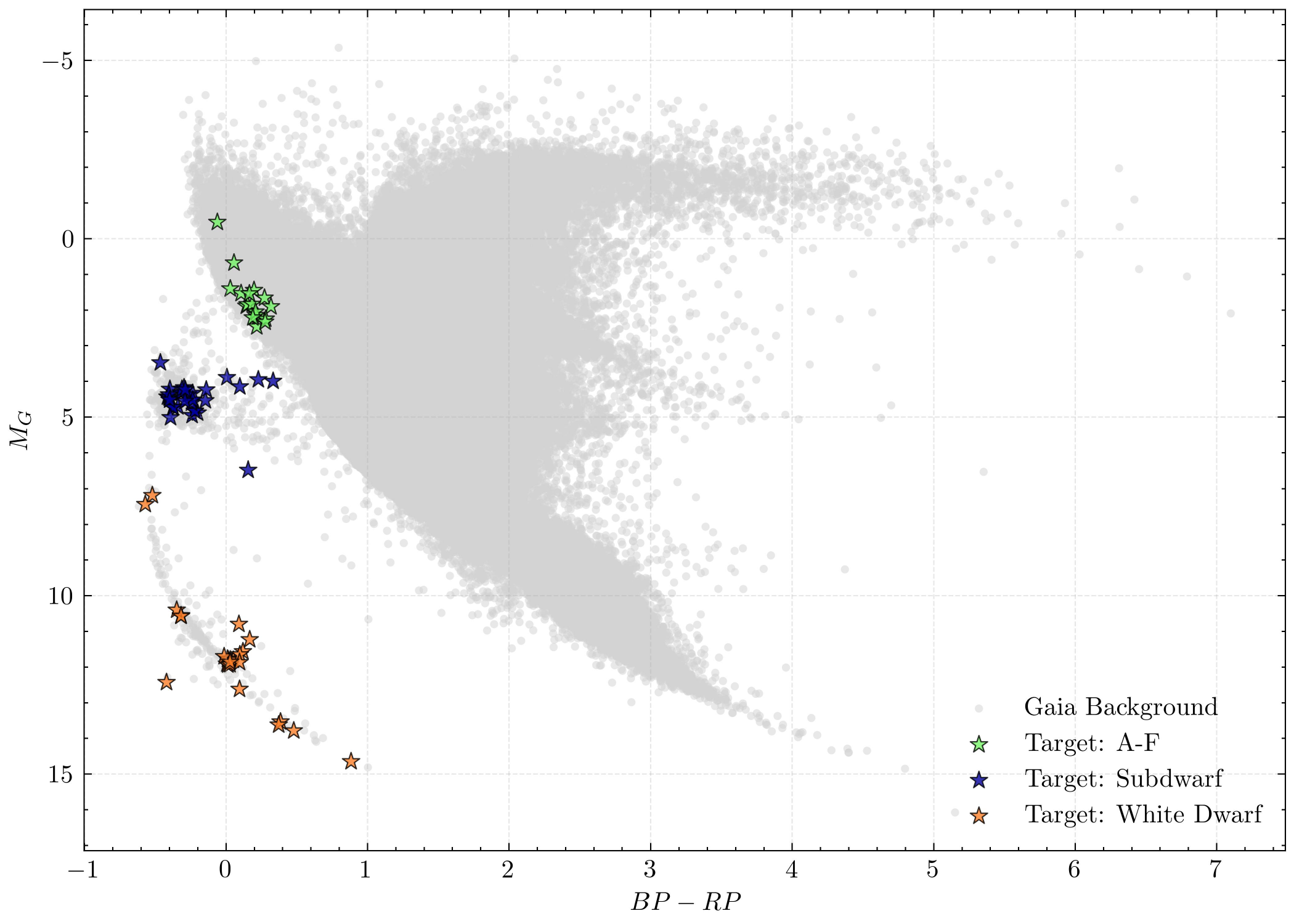}
\caption{Targets identified in this work shown on the Gaia $BP-RP$ versus $M_G$ 
color--magnitude diagram.
The grey background points represent a reference sample of field stars from Gaia DR3.
The colored symbols mark our newly discovered targets: orange stars indicate white dwarfs,
blue stars represent hot subdwarfs, and green stars denote A- and F-type variables. 
The targets occupy distinct regions corresponding to the white dwarf cooling sequence,
the extreme horizontal branch, and the upper main sequence, respectively.}
\label{fig:BP-RP vs GMAG}
\end{figure*}

\autoref{fig:BP-RP vs GMAG} plots the absolute G-band magnitude against the observed color index ($BP-RP$), 
presenting the locations of the targets identified in this work on the Gaia color--magnitude diagram (CMD).
The majority of the sample falls within the distinct domains characteristic of their respective classifications, 
confirming the overall consistency of our target selection. 
A background sample of field stars from Gaia DR3 is shown in grey to delineate the different stellar evolutionary stages, 
including the main sequence, the giant branch, and the white dwarf cooling sequence.
Most of the white dwarf candidates (denoted by orange stars) are tightly clustered along the canonical white dwarf cooling sequence, 
confirming their evolved status. 
The hot subdwarf stars (blue stars) are situated in the region intermediate 
between the main sequence and the white dwarf sequence ($M_G \sim 5$), 
a locus characteristic of the extreme horizontal branch (EHB) stars. 
The A- and F-type variables (green stars) are located near the upper main sequence and the turn-off point, 
consistent with the classical instability strip where $\delta$~Scuti and roAp stars are typically found. 
The clear separation of these populations in the CMD supports the robustness of our classification 
and reflects the distinct evolutionary stages associated 
with the observed pulsation properties.

\begin{figure}
\plotone{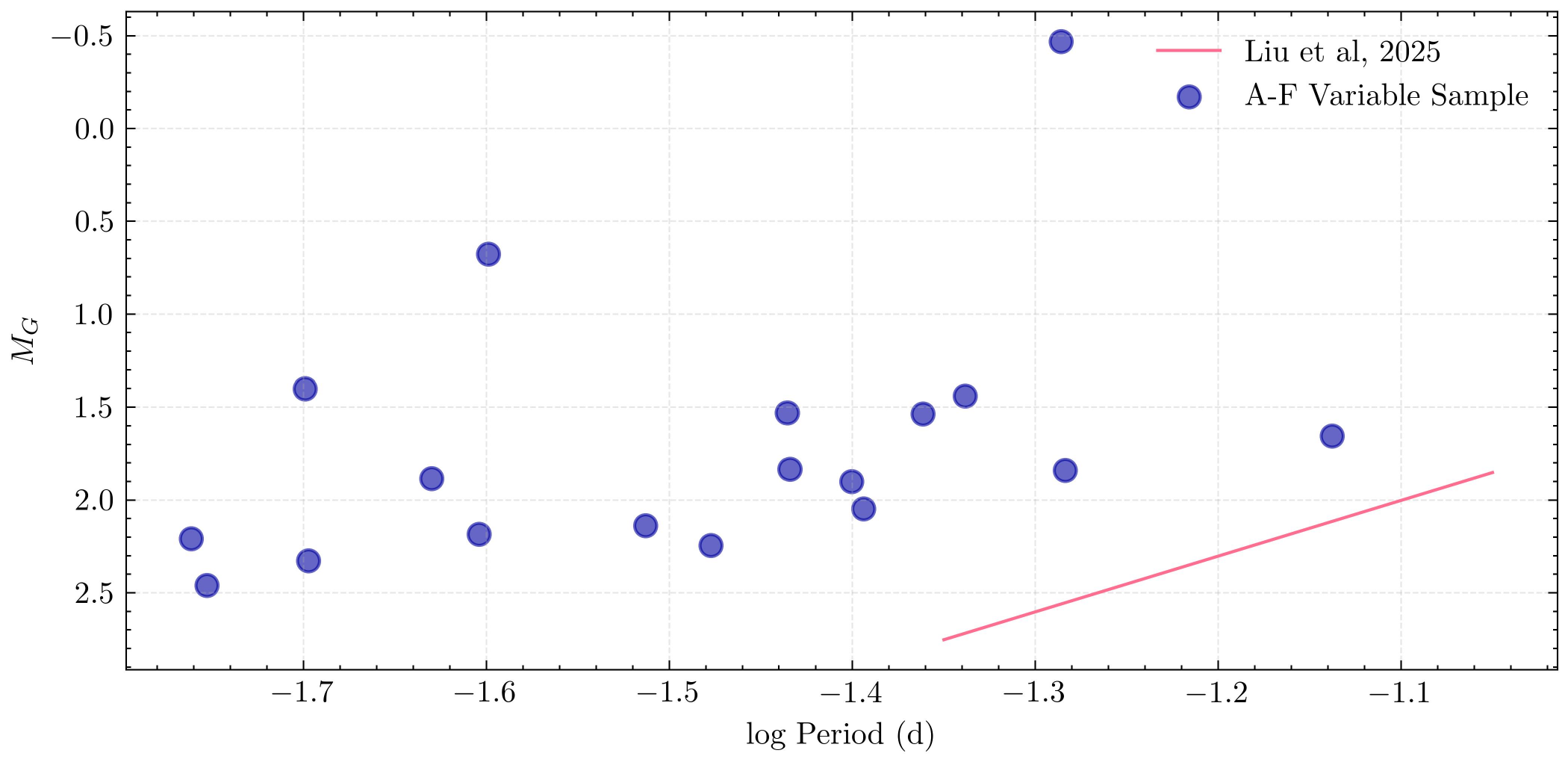}
\caption{
Logarithmic pulsation period versus Gaia absolute G magnitude for the A-F variable sample.
Each point represents one target.
The red solid line indicates the P-L relation of $\delta$~Scuti stars, following \cite{Liu2025}, 
with $M_G=a_1(\log{P}-\log{P_0})+a_2$, in which $a_1=-3.003$, $a_2=1.795$, $\log{P_0}=-1.031$. 
}
\label{fig:dsct_logP_GMAG}
\end{figure}

We further explore the relationship between pulsation period and luminosity for the A-F variable sample. 
\autoref{fig:dsct_logP_GMAG} plots the absolute Gaia G magnitude, as a proxy of luminosity, 
against the logarithmic dominant pulsation period $\log{P}$. 
Although the sample shows intrinsic scatter, 
a linear fit reveals a trend where stars with longer pulsation periods tend to be intrinsically brighter. 
The Pearson correlation coefficient is $r=-0.52$. 
The observed trend is consistent with expectations from the period--luminosity relation for $\delta$~Scuti stars, 
in which longer-period pulsations are generally associated with higher luminosities.
As shown in \autoref{fig:dsct_logP_GMAG}, our targets 
are systematically located at shorter periods compared to the prediction \citep{Liu2025}. 
This offset is expected, since our search focuses on rapidly oscillating targets, whose signals can be 
excited by higher-order overtones, which naturally have shorter periods than the fundamental radial mode at a given luminosity.

White dwarfs typically oscillate in the $f\gtrsim 60\ \mathrm{d}^{-1}$ 
frequency range with \textit{g}-modes, and rotational splitting can be observed in many cases (see examples in \citealt{Bognar2024}). 
In the limit of slow, rigid rotation, a mode of degree $l$ is split into $2l+1$ components with frequencies given by
\begin{equation}
f_{n,l,m} \simeq f_{n,l,0} + m (1 - C_{n,l}) \Omega ,
\end{equation}
where $m$ is the azimuthal order, $\Omega$ is the stellar rotation frequency, and $C_{n,l}\sim 1/l(l+1)$ is the Ledoux constant. 
Consequently, we can estimate the rotation period when rotational splitting is observed. 

\begin{figure*}[htbp]
\plotone{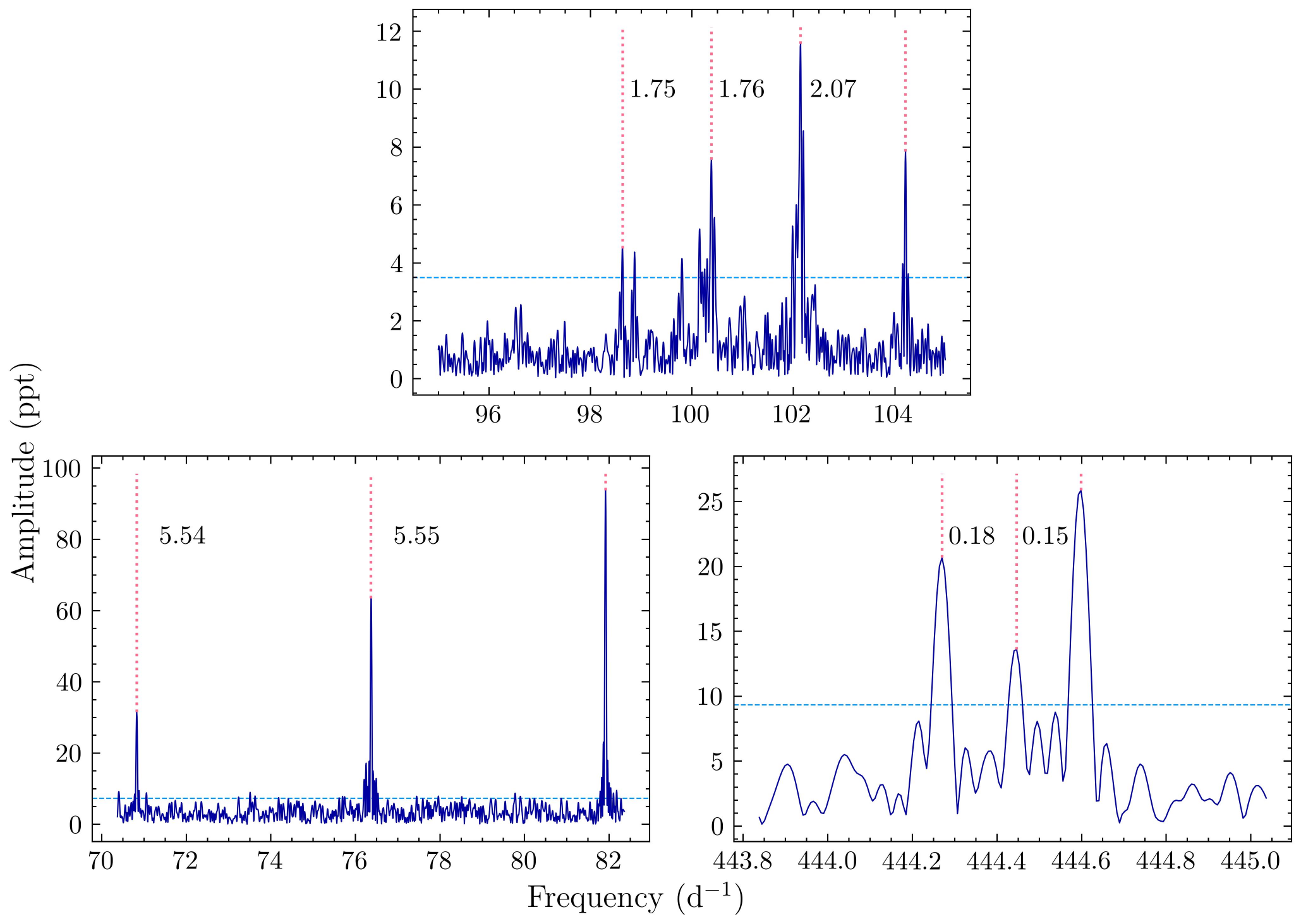}
\caption{
Amplitude spectra of one pulsating white dwarf and one hot subdwarf targets showing clear frequency multiplets that are likely caused by rotational splitting, 
and one CV showing a regular frequency spacing corresponding to the orbital frequency. 
Top panel: TIC 458484139, showing a multiplet structure centered at $100.4\ \mathrm{d}^{-1}$. The three dominant peaks form a clear triplet, with an additional nearby peak possibly representing a component of an incomplete $l=2$ multiplet. 
Bottom left: TIC 143496219, a very regular spacing $5.55\ \mathrm{d}^{-1}$ is detected, indicating the orbital frequency, 
modulating the spinning frequency of $76.369\ \mathrm{d}^{-1}$. 
Bottom right: TIC 426971746 reveals a triplet structure centered at $444.6\ \mathrm{d}^{-1}$. The side peaks are separated from the central peak by $0.18\,\mathrm{d}^{-1}$ and $0.15\,\mathrm{d}^{-1}$, respectively.
The horizontal blue dashed line indicates the adopted significance threshold at FAP$=0.0001$.
Vertical red dotted lines mark the detected peaks, and $\delta f$ (in $\mathrm{d}^{-1}$) is labeled between peaks.
}
\label{fig:rotational_splitting}
\end{figure*}

The top row of \autoref{fig:rotational_splitting} presents one pulsating white dwarf target, TIC 458484139, exhibiting clear frequency multiplets in the amplitude spectrum, 
which are naturally interpreted as the result of rotational splitting of non-radial pulsation modes. 
Three dominant peaks are clearly detected with nearly equal frequency spacing, 
strongly favoring an $l=1$ triplet interpretation. 
An additional peak at higher frequency is also present and may correspond to a single component of an $l=2$ multiplet. 
If this is the case, the remaining component of the quintuplet is either intrinsically low in amplitude or falls below the detection threshold. 
We therefore interpret the TIC 458484139 spectrum primarily as an $l=1$ rotationally split mode, while noting the possibility of an incomplete $l=2$ multiplet. 
Assuming the $l=1$ interpretation, the rotation period can be estimated as $0.285\ \mathrm{day}$.  

Our selection criterion of $f \geq 50~\mathrm{d}^{-1}$ aims at 
the domain of \textit{p}-mode and intermediate-mode subdwarf pulsators, 
distinct from the typical \textit{g}-mode regime of $f \lesssim 55~\mathrm{d}^{-1}$ 
(see \citealt{Reed2020} for an example of short period \textit{g}-mode oscillator).
Despite this, we detected \textit{g}-mode signals in TIC~25836205 and TIC~278271659, 
indicating these are hybrid pulsators with concomitant high-frequency oscillations. 
TIC~860899761 and TIC 452712793 represent two exceptional cases situated at the transition boundary; 
while they can be classified as \textit{g}-mode pulsators, 
the oscillations are confined to the extreme upper limit of the \textit{g}-mode frequency range, 
in TIC 860899761, the dominant signal is detected at $54.976~\mathrm{d}^{-1}$, 
while in TIC 452712793, a dominant signal is detected at $50.712\ \mathrm{d}^{-1}$, 
accompanied by a signal at $56.946\ \mathrm{d}^{-1}$. 
TIC 30339103 is another special case, as detected in a binary system. 
With all orbital frequencies and its harmonics removed in the prewhitening procedure, 
several peaks in the intermediate-mode are detected. 

The frequency spacing was found in two cases of the subdwarf candidates as well. 
One of these is TIC~143496219, in which 
we identified a very regular frequency spacing of $~5.55\ \mathrm{d}^{-1}$, 
as the bottom left panel of~\autoref{fig:rotational_splitting} shows.
However, the high amplitude is somewhat anomalous for a pulsating subdwarf. 
As we inspected the periodogram and light curve of the target, it became clear that 
TIC~143496219 is a cataclysmic variable (CV) with subclass known as intermediate polar (IP) \citep{GaiaDR3p4Variability}. 
The frequency peak at $81.916\ \mathrm{d}^{-1}$ is identified as the spinning frequency, and 
modulated by an orbital frequency of $5.55\ \mathrm{d}^{-1}$. 
In addition to the dominant coherent signals discussed above, 
several low-amplitude peaks are present in the power spectrum at frequencies below $40\ \mathrm{d}^{-1}$. 
These signals are not phase-coherent over the duration of the observations 
and do not exhibit simple commensurabilities with either the orbital or spin frequencies. 
We therefore attribute them to stochastic or quasi-periodic variability 
associated with accretion processes, and do not consider them further in this work. 
TIC~426971746 in the bottom right panel of~\autoref{fig:rotational_splitting} 
is found to have a much smaller frequency spacing, 
with an average spacing of $\sim0.165\ \mathrm{d}^{-1}$. 
Considering a rotational splitting interpretation, a rotation period of 
$6.060\ \mathrm{d}$ can be inferred, since the Ledoux constant for \textit{p}-mode is close to zero. 
This period is relatively short compared to those of single pulsating hot subdwarf stars, 
as single pulsating subdwarf B stars are slow rotators, 
typically with a rotation period of several tens of days 
(e.g. \citealt{Baran2012, Telting2012, Ostensen2014, Foster2015, Charpinet2018}). 
However, subdwarf B stars in close binary systems can have much shorter rotation periods due to
tidal interactions, ranging from $\sim1\ \mathrm{h}$ to $\sim10\ \mathrm{d}$ \citep{Geier2010, Preece2018}. 
Therefore, we note that TIC~426971746 is a candidate close binary system with an oscillating subdwarf. 

\begin{figure}
\plotone{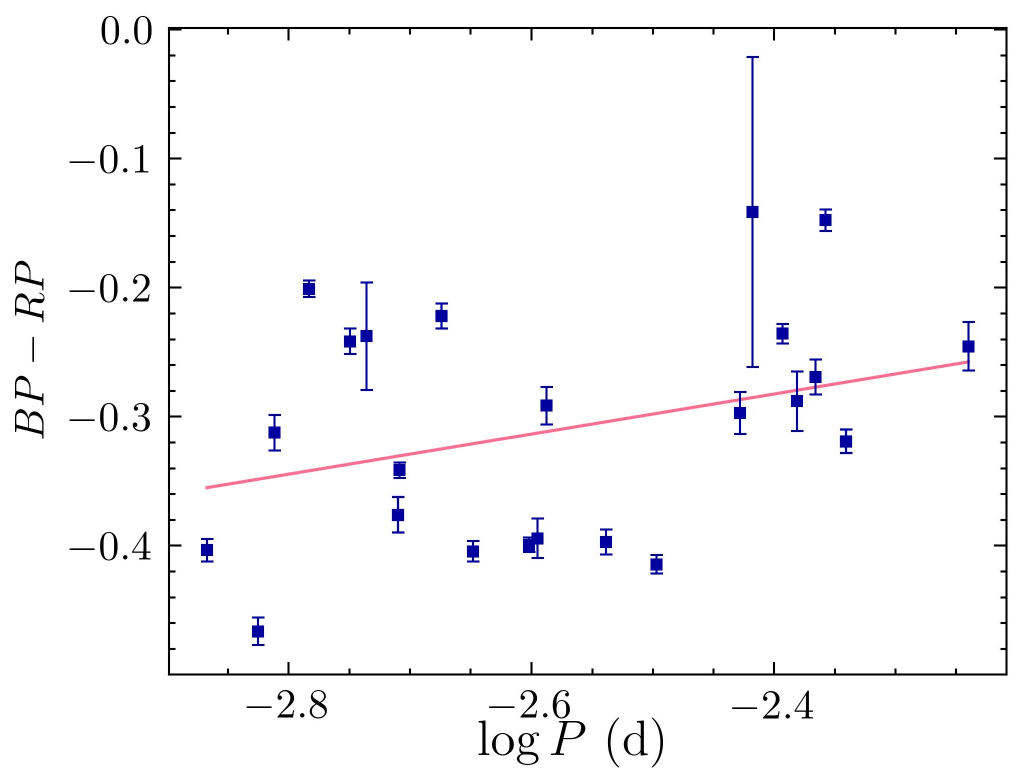}
\caption{
The Gaia $BP-RP$ color index plotted against the logarithm of the period ($P$ in days) 
for the \textit{p}-mode hot subdwarf sample. 
The blue squares with error bars represent the observed data points. 
The solid red line indicates the best-fitting linear regression to the data. 
Note that targets with $\log P<-3.0$ and $\log P>-2.0$ have been excluded from this plot 
to highlight the period-color trend in the primary population.
}
\label{fig:sd_logp_vs_color}
\end{figure}

\autoref{fig:sd_logp_vs_color} displays the Gaia $BP-RP$ color index as a function of the logarithmic pulsation period ($\log P$) 
for our \textit{p}-mode subdwarf target sample. To ensure a robust analysis, 
we excluded three outliers at the extreme periods of the sample ($\log P<-3.0$ or $\log P>-2.0$) from the dataset.
Despite the intrinsic scatter observed in the data, 
a discernible positive correlation is evident, 
as illustrated by the linear regression fit (solid red line). 
This trend indicates that \textit{p}-mode subdwarfs with longer pulsation periods tend to exhibit redder colors. 

\section{Discussion}
\label{sec:discussion}

The large pixel scale of TESS ($21''$) makes pixel-level vetting essential, 
especially for faint targets whose apertures can be affected by neighboring sources, unresolved blends, 
or background-dominated flux extraction. 
In our sample, we identified three representative contamination scenarios: 
(i) flux leakage from nearby bright variable stars, 
(ii) unresolved optical doubles within a single TESS pixel, 
and (iii) background-dominated faint targets.

One notable example is TIC 711886697 ($\alpha_{\rm J2000} = 06^{\rm h} 13^{\rm m} 21\fs390$, $\delta_{\rm J2000} = -01\degr 04\arcmin 11\farcs16$), 
a faint white dwarf candidate ($G_{mag} \approx 19$). 
We detected a high-amplitude signal ($\sim 200\ \mathrm{ppt}$) in the initial analysis. 
Given the suspiciously high amplitude for a pulsating white dwarf target, 
we inspected the Target Pixel File (TPF) and the 2-minute cadence light curve 
of a nearby bright star, TIC 241986712. 
Analysis of the Sector 33 data for TIC 241986712 revealed variability at frequencies 
identical to those observed in the target. 
Furthermore, a custom aperture light curve of TIC 241986712 extracted from the Sector 87 TPF 
exhibits the same variability pattern. 
Consequently, we attribute the variability detected in TIC 711886697 to contamination from TIC 241986712. 

We highlight a noteworthy pair of targets, TIC 267166357 and TIC 267166358. 
Upon detecting signals at identical frequencies in both stars, we inspected the Sector 93 TPF data. 
We found that this optical double is unresolved, falling within one single TESS pixel, 
making it impossible to photometrically distinguish the source of variability. 
We have retained the signal in our final catalog, attributing it to TIC 267166357. 

The third scenario can affect a broader class of faint targets, 
for which the extracted aperture flux may be dominated by background emission or residual systematics. 
In such cases, apparently significant high-frequency peaks can be difficult to interpret without careful TPF inspection. 
TIC 1255689804 provides a detailed example of this situation.
TIC 1255689804 ($\alpha_{\rm J2000} = 16^{\rm h} 10^{\rm m} 22\fs161$, $\delta_{\rm J2000} = -31\degr 04\arcmin 19\farcs53$) is a faint white dwarf target ($G_{mag}=19.3$) 
where we initially detected three signals at frequencies of $\sim 1155$, $809$, 
and $1596\ \mathrm{d}^{-1}$, all with amplitudes of $\sim 200\ \mathrm{ppt}$. 
We examined the Sector 91 TPF and extracted custom aperture light curves for two nearby bright stars, 
Gaia DR2 6036767132312273920 at $\alpha_{\rm J2000} = 16^{\rm h} 10^{\rm m} 23\fs613$, $\delta_{\rm J2000} = -31\degr 04\arcmin 39\farcs700$ 
and Gaia DR2 6036767132312283648 at $\alpha_{\rm J2000} = 16^{\rm h} 10^{\rm m} 23\fs991$, $\delta_{\rm J2000} = -31\degr 03\arcmin 56\farcs797$ with $G_{mag}=11.9$ and $13.7$, respectively. 
While no variability was found in the neighbors, subsequent analysis of TIC 1255689804 
using both pipeline-defined aperture and custom apertures failed to recover the signals in a robust or reproducible manner. 
An inspection of the \texttt{sap\_flux} in the light-curve products revealed
that the flux within the aperture is dominated by background emission: 
despite the removal of $5\sigma$ outliers, the flux values ranged from a maximum of $1300.50\,e^-\,\mathrm{s}^{-1}$ 
to a minimum of $859.49\,e^-\,\mathrm{s}^{-1}$. 
As the expected flux for this target can be estimated to 
$\lesssim 10\,e^-\,\mathrm{s}^{-1}$ based on its Gaia magnitude, this flux value suggests that 
the \texttt{sap\_flux} is dominated by background flux, and when processed through SPOC pipeline, 
the resulting \texttt{pdcsap\_flux} light curve is consistent with noise-dominated residuals, 
with the signals detected therein most likely dominated by 
residual instrumental or background-related systematics rather than intrinsic stellar variability. 
We consequently exclude this target from our final results. 
This example shows how background-dominated extraction can mimic high-amplitude rapid variability. 
While detailed TPF inspection was required to diagnose TIC 1255689804, 
many other faint targets ($G_{mag} \gtrsim 18$) in our initial sample showed more readily identifiable 
unphysical light-curve characteristics. 
We therefore provide a representative list of such rejected faint targets from Sector 84 in \autoref{table:false_target_examples}. 
These examples illustrate the following empirical indicators of severe contamination, 
background-dominated extraction, or instrumental/systematic effects introduced during pipeline processing:
\begin{enumerate}
    \item Extremely high-amplitude signals (e.g., hundreds of parts per thousand) that are uncharacteristic for the expected stellar type;
    \item A large number of significant peaks at unusually high frequencies (e.g., hundreds to thousands of $\mathrm{d}^{-1}$);
    \item Strong flux variations in the light curves, with relative flux fluctuations of $\gtrsim4$ times (or even higher than) the median flux level.
\end{enumerate}
Targets exhibiting multiple such features are highly likely to be false positives 
and should be carefully evaluated or excluded to ensure catalog reliability.

\section{Summary}
\label{sec:summary}
In this work, we have conducted a systematic search for high-frequency stellar variability 
using the 20-second cadence data from TESS Cycle 7 (Sectors 84--96). 
By analyzing approximately $3.9\times10^{4}$ light curves, 
we targeted the regime of rapid oscillations ($f\geq50\ \mathrm{d}^{-1}$), 
a domain that allows for the probing of short-timescale physics in compact pulsators 
and high-overtone main-sequence variables. 
Through a rigorous frequency extraction pipeline, which incorporates window-function filtering, 
iterative prewhitening, and a strict significance threshold of $FAP\leq10^{-4}$, 
we established a homogeneous catalog of rapid oscillators. 
A crucial component of this study was the validation of signals against instrumental artifacts 
and background contamination. 
The large pixel scale of TESS necessitates careful inspection of Target Pixel Files, 
particularly for faint compact objects like white dwarfs. 
We demonstrated that high-amplitude signals in faint sources can often be attributed to contamination 
from bright nearby variables or background systematics, 
underscoring the importance of custom aperture analysis in high-frequency TESS surveys. 
Our final sample consists of 73 newly identified rapid variables, 
comprising 24 pulsating white dwarfs, 31 hot subdwarf stars, and 18 A--F type stars. 
The positions of these targets on the Gaia color-magnitude diagram validate their evolutionary classifications, 
spanning from the main sequence to the white dwarf cooling track. 
Among the key findings, we detected clear frequency multiplets in one white dwarf and one hot subdwarf,  
(TIC 458484139 and TIC 426971746) consistent with rotational splitting, 
enabling estimates of their rotation periods. 
We also characterized the diverse variability of hot subdwarfs, 
identifying pure \textit{p}-mode pulsators, hybrid candidates, and binary systems 
such as TIC 426971746, where tidal effects likely influence the pulsation spectrum. 
Furthermore, the A-F sample exhibits a $\delta$~Scuti period-luminosity trend, 
though shifted toward shorter periods as expected for the high-frequency regime targeted here. 
Looking ahead, this uniformly processed sample provides a robust foundation for ensemble asteroseismology of compact stars. 
As TESS continues its survey, expanding this baseline to future sectors will improve statistical constraints on the instability strips of evolved stars. 
However, photometric detection is only the first step; dedicated ground-based spectroscopic follow-up is essential to precisely determine atmospheric parameters ($T_{eff}$ and $\log g$) and chemical abundances for these faint targets. 
Such multi-messenger datasets will be critical for secure mode identification, 
particularly for resolving the ambiguity between dipole and quadrupole modes in rotationally split multiplets, 
and for ultimately refining our models of stellar interiors and angular momentum evolution in the late stages of stellar life.

\section*{Acknowledgements} 
We thank the anonymous referee for their constructive comments and suggestions, which have significantly improved the quality of this paper. This work was supported by the National Natural Science Foundation of China (NSFC) through grants 12322306, 12373028, 12373037, 12233009, 12133002, 12173047. We also acknowledge support from the National Key Research and Development Program of China through grant 2022YFF0503404. X. C. and S. W. acknowledge support from the Youth Innovation Promotion Association of the CAS (grant Nos. 2022055 and 2023065). This paper includes data collected with the TESS mission, obtained from the MAST data archive at the Space Telescope Science Institute (STScI). Funding for the TESS mission is provided by the NASA Explorer Program. STScI is operated by the Association of Universities for Research in Astronomy, Inc., under NASA contract NAS 5-26555.

\FloatBarrier
\onecolumngrid
\appendix
\section{Result Tables}
\label{sec:tables}

\begin{longdeluxetable}{l c c c l}
    \tablewidth{0pt}
    \tablecaption{White dwarf oscillators in this work. \label{table:1}}
    \tablehead{
        \colhead{TIC} & 
        \colhead{Sector} & 
        \colhead{Frequency ($\mathrm{d}^{-1}$)} & 
        \colhead{Amplitude ($\mathrm{ppt}$)} & 
        \colhead{Note}}
    \startdata
        26517742 & 84 &  138.820 &  23.971 & \\
        & &  253.796 &  19.858 & \\
        259140053 & 84 &  238.759 &  13.051 & \\
        458484139 & 84 &  102.141 &  11.568 &  Sidelobes are emitted. \\
        & &  159.599 &  10.088 & \\
        & &  104.209 &  7.884 & \\
        & &  100.383 &  7.617 & \\
        & &  107.488 &  6.563 & \\
        & &  214.307 &  6.000 & \\
        & &  98.625 &  4.368 & \\
        2042782309 & 84 &  136.442 &  19.984 & \\
        167394689 & 87 &  98.358 &  15.635 & \\
        333432673 & 87 &  189.641 &  17.675 & \\
        & &  174.148 &  14.612 & \\
        & &  218.984 &  8.488 & \\
        685214650 & 87 &  787.464 &  69.118 & \\
        77499896 & 91 &  151.606 &  74.727 & \\
        & &  176.788 &  39.420 & \\
        & &  113.942 &  29.605 & \\
        & &  328.400 &  29.127 & \\
        225798235 & 91 &  73.066 &  13.100 & \\
        242073510 & 92 &  135.487 &  39.457 & \\
        442532281 & 92 &  190.043 &  14.758 & \\
        267166357 & 93 &  119.051 &  70.776 & \\ 
        & &  238.102 &  5.721 & \\ 
        451533898 & 93 &  84.386 &  3.328 & \\
        & &  94.526 &  2.401 & \\
        & &  185.049 &  2.312 & \\
        1508985146 & 93 &  108.990 &  31.867 & \\
        1522550276 & 93 &  129.513 &  75.293 & \\
        389348640 & 94 &  102.117 &  7.324 &  Sidelobes are emitted. \\
        & &  97.512 &  5.237 & \\
        & &  122.926 &  4.986 & \\
        & &  138.266 &  4.966 & \\
        & &  106.934 &  2.862 & \\
        & &  235.316 &  2.364 & \\
        & &  92.981 &  2.176 & \\
        & &  115.639 &  2.176 & \\
        & &  9.592 &  1.468 & \\
        & &  88.863 &  1.311 & \\
        1990094034 & 94 &  336.375 &  10.454 & \\
        & &  307.488 &  9.330 & \\
        53851007 & 95 &  331.953 &  6.614 & \\
        1989122424 & 95 &  214.148 &  27.235 & \\
        & &  213.133 &  17.963 & \\
        2051988598 & 96 &  117.543 &  21.580 & \\
        267885913 & 84 &  182.908 &  17.335 & \\
        628720291 & 85 &  211.527 &  58.702 & \\
        641489462 & 85 &  139.112 &  21.727 & \\
        313144249 & 86 &  33.175 &  36.552 & \\
        & &  66.350 &  13.468 & \\
    \enddata
\end{longdeluxetable}

\begin{longdeluxetable}{l c c c l}
    \tablewidth{0pt}
    \tablecaption{Subdwarf oscillators in this work. \label{table:2}}
    \tablehead{
        \colhead{TIC} & 
        \colhead{Sector} & 
        \colhead{Frequency ($\mathrm{d}^{-1}$)} & 
        \colhead{Amplitude ($\mathrm{ppt}$)} & 
        \colhead{Note}}
    \startdata
        129218058 & 84 &  607.001 &  12.943 & \\
        237806064 & 87 &  561.478 &  13.007 & \\
        156991809 & 88 &  345.833 &  10.847 & \\
        & &  346.740 &  4.932 & \\
        25836205 & 89 &  314.063 &  13.218 &  \\ 
        & &  311.752 & 8.003 & \\
        & &  312.543 & 2.267 & \\
        & &  33.972 &  1.858 & \\
        & &  28.206 &  1.647 & \\
        & &  417.172 &  1.412 & \\
        & &  22.567 &  1.305 & \\
        & &  42.810 &  1.247 & \\
        & &  30.870 &  0.895 & \\
        & &  24.045 &  0.876 & \\
        & &  30.657 &  0.760 & \\
        & &  29.805 &  0.651 & \\
        143496219 & 89 &  81.916 &  93.766 & \shortstack[l]{CV white dwarf spinning frequency $f_{\text{spin}}$.\\Lower-amplitude, non-coherent signals are emitted.}  \\
        & &  76.369 &  63.393 & $f_{\text{spin}} - f_{\text{orb}}$ \\
        & &  70.827 &  31.644 & $f_{\text{spin}} - 2f_{\text{orb}}$ \\
        & &  152.743 &  22.456 & $n=2$ harmonic of $f_{\text{spin}} - f_{\text{orb}}$. \\
        650355986 & 89 &  512.536 &  7.459 & \\
        807661828 & 89 &  472.068 &  24.835 & \\
        5816844 & 90 &  623.327 &  2.752 & \\
        61541623 & 90 &  736.880 &  4.177 & \\
        & &  626.259 &  3.049 & \\
        345178165 & 90 &  511.058 &  7.583 & \\
        452712793 & 90 &  50.712 &  0.033 & \\
        & &  56.946 &  0.030 & \\
        860899761 & 90 &  54.976 &  29.701 & \\
        & &  45.201 &  13.120 & \\
        867256079 & 90 &  173.692 &  68.500 & \\
        399540269 & 91 &  84.884 &  37.966 & \\
        & &  85.651 &  26.255 & \\
        & &  88.017 &  9.103 & \\
        & &  170.529 &  6.276 & \\
        30339103 & 92 &  6.934 &  408.290 &  \shortstack[l]{Binary orbital frequency. Sidelobes and\\orbital harmonics are emitted.} \\
        & &  267.962 &  29.765 & \\
        & &  264.203 &  11.399 & \\
        & &  269.757 &  10.759 & \\
        & &  245.383 &  7.471 & \\
        1571064278 & 92 &  544.404 &  96.575 & \\
        22579708 & 93 &  232.299 &  34.784 & \\
        265185984 & 93 &  227.786 &  8.844 & Sidelobes are emitted. \\ 
        278271659 & 93 &  247.318 &  16.878 & Sidelobes are emitted. \\
        & &  249.538 &  15.341 & \\
        & &  26.636 &  3.009 & \\
        & &  16.472 &  2.607 & \\
        & &  17.987 &  1.736 & \\
        & &  245.845 &  1.684 & \\
        & &  29.051 &  1.565 & \\
        294099979 & 93 &  219.203 &  7.959 & \\
        323572215 & 93 &  1333.285 &  0.964 &  Sidelobes are emitted. \\ 
        403003556 & 93 &  386.953 &  18.406 & \\
        & &  381.746 &  16.223 & \\
        426971746 & 93 &  444.599 &  25.862 &  \\ 
        & & 444.270 & 20.118 & \\
        & & 444.440 & 13.383 & \\
        & & 446.259 & 12.588 & \\
        & & 446.064 & 10.612 & \\
        1697850501 & 93 &  261.807 &  7.966 & \\
        1704610980 & 93 &  240.487 &  34.579 & \\
        230973606 & 94 &  648.491 &  6.845 & \\
        278403480 & 94 &  809.422 &  5.240 & \\
        469791892 & 94 &  400.341 &  1.868 & \\
        631694614 & 96 &  668.942 &  21.248 & \\
        390914725 & 92 &  393.881 &  17.151 & \\
        1318844564 & 93 &  78.437 &  111.273 & \\
    \enddata
\end{longdeluxetable}

\begin{longdeluxetable}{lccc}
    \tablecaption{A-F type variables in this work. Only dominant frequency and amplitude 
    of each target are listed here. The full table is available in the machine-readable format.\label{table:A-F variables}}
    \tablehead{
        \colhead{TIC} & \colhead{Sector} &
        \colhead{Frequency($\mathrm{d}^{-1}$)} & \colhead{Amplitude($\mathrm{ppt}$)}}
    \startdata
        16919805 & 84 & 40.189 & 1.197 \\
        219751969 & 84 & 39.708 & 0.175 \\
        243188395 & 84 & 42.646 & 1.901 \\
        426960606 & 84 & 19.215 & 0.812 \\
        440658751 & 84 & 27.172 & 0.14 \\
        367160856 & 85 & 32.585 & 0.771 \\
        418223905 & 85 & 21.800 & 0.229 \\
        392878513 & 86 & 13.735 & 2.106 \\
        42569616 & 87 & 27.269 & 0.222 \\
        412682111 & 87 & 22.993 & 0.614 \\
        145713666 & 89 & 24.769 & 2.149 \\
        24373893 & 92 & 25.152 & 3.021 \\
        135485021 & 92 & 30.012 & 2.487 \\
        433218066 & 92 & 50.030 & 0.039 \\
        361861044 & 93 & 57.731 & 8.034 \\
        202431888 & 84 & 19.319 & 0.173 \\
        253877224 & 94 & 49.824 & 2.538 \\
        254068584 & 94 & 52.549 & 3.858 \\
    \enddata
\end{longdeluxetable}

\begin{longdeluxetable}{ccl}
    \tablecaption{Representative Examples of Rejected Faint Targets (Sector 84) \label{table:false_target_examples}}
    \tablehead{
    \colhead{TIC ID} & \colhead{Gaia $G$ mag} & \colhead{Primary Anomaly Flag / Note}
    }
    \startdata
        601142149 & 19.00 &  \shortstack[l]{Unreliable $FAP$ estimation due to a contaminated or noise-dominated light curve, \\resulting in a large number of spurious peaks. }\\
        1401004611 & 18.14 & Same anomaly pattern as TIC 601142149. \\
        1551589054 & 18.22 & Same anomaly pattern as TIC 601142149.  \\
        2021198496 & 19.31 & Same anomaly pattern as TIC 601142149.  \\
        620537762 & 19.38 &  \shortstack[l]{A large number of peaks found in the extremely high-frequency domain ($\sim1000\to\sim2000\ \mathrm{d}^{-1}$) \\with large amplitudes ($\sim200\ \mathrm{ppt}$). }\\
        1271281336 & 19.10 & The only peak is found at the upper limit of the calculated frequency range ($\sim2160\ \mathrm{d}^{-1}$).  \\
        1884488768 & 19.08 & Same anomaly pattern as TIC 1271281336.  \\
        2023424609 & 18.79 & \shortstack[l]{A clearly noise-dominated light curve in which a variability range of \\$\pm \sim 40 $ times the median flux level is detected. }\\
        2043603493 & 19.13 & Same anomaly pattern as TIC 2023424609.  \\
        603424090 & 19.49 &  \shortstack[l]{Same anomaly pattern as TIC 2023424609, except that the flux fluctuates by \\ $\pm\sim100$ times the median flux level. }\\
    \enddata
\end{longdeluxetable}

\bibliographystyle{aasjournalv7}
\bibliography{references}

\end{document}